# Psychological Frameworks for Persuasive Information and Communications Technologies

Joseph J.P. Simons
*Social and Cognitive Computing Department*
*Institute of High Performance Computing, Singapore*

*When developing devices to encourage positive change in users, social psychology can offer useful conceptual resources. This article outlines three major theories from the discipline and discusses their implications for designing persuasive technologies.*

Developers working on devices or systems that aim to make users happier, healthier, or more motivated would do well to talk to a social psychologist. (The experience of trying to influence human behavior might also drive developers to seek out a psychotherapist, but that's a separate matter.) Since the 1930s, researchers have been conducting rigorous, quantitative studies probing the causal processes underlying social influence, attitude change, and motivation. The discipline provides a fertile resource for understanding the drivers of people's thoughts, feelings, and behaviors.

Information and communication technologies (ICTs) that interface well with human psychological mechanisms are more engaging, exciting, and motivating for users. A good example is the Zombies, Run! game (www.zombiesrungame.com), a fitness app that incorporates standard features of a running app — such as recording performance —with a layer of narrative and game-like elements that occur both during and between running sessions. During the run, users are on a mission, and they find items, experience zombie attacks, and discover more of the backstory. Between runs, they can make decisions about how to deploy in-game resources. This structure packs two levels of psychological sophistication: the feedback on performance alone can be motivating, but creating a larger immersive experience also taps into users' intrinsic motivation.

How can developers achieve such design sophistication without a PhD in psychology? Here, I present three frameworks for use in the development of pervasive technologies intended to foster positive change:

- The *theory of planned behavior* provides an account of the factors influencing conscious intentions. It's particularly useful for understanding why people aren't always resolved to engage in positive behaviors, even when they recognize the benefits of doing so.
- *Self-determination theory* addresses when people will be internally motivated to engage in a behavior. It's particularly suited for understanding how deep psychological change (as opposed to more superficial compliance) can be induced.
- *Control theory* provides an account of the process of goal pursuit. It's particularly useful for understanding why positive intentions don't always translate into action.

My goal is to provide developers with a flexible and generative conceptual toolbox for bringing psychology to bear on ICT design.

**Psychology and ICT: an overview**

Just as different tools are relevant for different aspects of a task, different theories can be useful for conceptualizing different parts of a problem. Similarly, just as tools can be useful for a wider or narrower range of tasks, theories can be relevant for solving a broader or narrower range of problems. The three theories discussed here constitute foundational tools that complement each other and can provide insight across a broad range of cases. This article therefore provides "conceptual frameworks" in the terminology of Eric Hekler and his colleagues.[1] The aim isn't to provide a fine-grained list of particular techniques to use or a universal theory of all behavior change, but rather to equip developers with practical working models of how devices affect users.

**Why these theories?**

Psychology has generated a wide range of theories and techniques for influencing behavior. A recent taxonomy has identified more than 90 different intervention techniques in the health domain alone.[2] Here, I highlight three theories, selected on the basis of practical utility to developers. In part, this usefulness is due to a solid history of application (as references cited later attest). However, perhaps the most important feature is their conceptual fertility. The theories presented abstract away from the details of specific application cases without seeking to provide a unitary account of all mental functioning. Hence, they provide a set of complementary perspectives, each of which can be applied to gain insight into new application cases—a particularly useful feature given how fast pervasive technology evolves.

Other psychological approaches can be relevant to developers. For example, behavioral economics and the psychology of judgment probe the heuristic rules-of-thumb by which people make sense of the world (such as using mental availability to judge probability). By examining these mental quirks, researchers hope to uncover why people sometimes make seemingly "irrational" decisions. A popular overview of basic research is provided by Daniel Kahneman,[3] and Richard Thaler and Cass Sunstein[4] apply these ideas to changing behavior via noncoercive "nudges."

A second approach is positive psychology, which is the study of optimal human experiences (including the study of well-being, the sense of meaning, and flourishing). An influential early example is Mihaly Csikszentmihalyi's study of flow experience,[5] the state of intrinsic interest and absorption stemming from a calibration of situational demands to individual ability. Jane McGonigal[6] has profitably applied ideas from this area to designing games that promote growth in players. Clearly, these approaches can offer insights for specific design challenges. However, in keeping with the toolkit metaphor, I focus here on three theories that, taken collectively, offer a fruitful starting point for conceptualizing a wide range of use cases.

**Psychology in ubicomp research**

Many ubicomp researchers already make good use of psychological theory in their work. In some cases, this involves using particular theories to inform the design of devices. For example, the design of the UbiFit system was guided by the theory of goal-setting.[7] The system motivated participants to exercise by setting specific goals rather than by exhorting them to do their best, allowing them to set their own goals rather than assigning general expert advice, and requiring goals to be set such that it was unambiguous whether or not they were met.

Other researchers have presented integrated models for changing behavior with technology. For example, the Fogg Behavior Change model identifies three main factors required for behavior change: motivation to engage in the behavior, the ability to do so, and specific behavior triggers.[8]

In other cases, researchers discuss the role of theory in the development process. Hekler and his colleagues describe how using theories can facilitate decisions about functionality, assessments of efficacy, and targeting of the most amenable user groups.[1] Predrag Klasnja, Sunny Consolvo, and Wanda Pratt[9] elaborate on the role of theory in evaluation, arguing that having a hypothesized mechanism of action opens up more tailored ways of assessing the effectiveness of new technologies. My article here is intended to contribute to this ongoing interdisciplinary dialogue by providing a practical primer on some fertile conceptual frameworks.

**Theory and the development process**

The interaction between technology and psychology is a two-way street. As I noted earlier, researchers have identified many ways in which psychological theories can guide the development process.[1,7] However, ubicomp devices are an especially potent medium for refining psychological theories. As Hekler and his colleagues discuss, these devices often have several advantages over traditional self-report research methods.[1] They offer more accurate measurements, reduce user burden (and hence allow longer studies with more frequent assessment), and unlock big data research strategies. The EmotionSense platform[10] provides an example of this in action; through collaboration between computer scientists and psychologists, it provides a mobile phone application for the measurement of emotion in everyday life. Although the emphasis of my work here is on the resources psychology can offer developers, it isn't meant to deny the importance of new technologies for theory testing and refinement.

My discussion of the three theories will focus on the relevance of each theory for ICT design. The goal of these discussions is to show how the psychological constructs postulated by these theories can provide a working model of how device features affect users. To serve this purpose, I emphasize concrete examples of how some commonly used device features can be understood within the terms of these theories. I also discuss the utility of the theories in highlighting potential "boomerang effects."

These are cases in which well-intentioned attempts to foster positive change have the opposite effect, actually increasing undesirable behaviors.

## The Theory of Planned Behavior

The theory of planned behavior seeks to understand deliberative behaviors, such as resolving to recycle more or quit smoking. It provides a set of concepts for thinking about conscious, reflective decisions, as opposed to impulsive, spontaneous actions. A detailed description of the theory is given by Icek Ajzen,[11] and Christopher Armitage and Mark Connor provide a meta-analysis of the theory's predictive efficacy over 185 independent empirical studies.[12]

### Attitudes, norms, and control

The theory postulates three precursors to intentions:
- *attitudes*—the individual's beliefs *about* the consequences of engaging in the behavior;
- *subjective norms*—the individual's beliefs about whether people they care about would want them to engage in the behavior; and
- *perceived behavioral control*—the degree to which the individual perceives the behavior as under his or her control.

Figure 1 shows the relationships between these constructs: all three determine intention to engage in the behavior. People are more likely to intend to do X if they think it will lead to good outcomes, that other people would want them to do it, and that it's under their control.

Actually engaging in the behavior is determined by intentions but also by perceived behavioral control. People are more likely to actually do X when they intend to do so and perceive X to be under their control.

To illustrate these points, take the example of recycling behavior. John, Mary, and David all fail to recycle, but for different reasons:
- John isn't convinced that recycling will really have much benefit on the environment; he thus views it as a waste of time.
- Mary's friends all joke that recycling is for granola-munching hippies.
- David doesn't think he has time in his schedule and anticipates he would just end up forgetting.

In this example, John has negative attitudes, Mary has negative subjective norms, and David has low perceived control.

### Intentions require more than just attitudes

In terms of implications for ICT design, the theory of planned behavior highlights that people's intentions are often influenced by factors beyond their own attitudes. It's intuitive to emphasize a person's opinions as a cause of their behavior. For example, if we want to make people recycle more, we might try to convince them that recycling is important. However, the theory of planned behavior makes explicit two additional precursors to action: subjective norms and perceived control. Devices that aim to change people's conscious intentions can, therefore, act on these constructs.

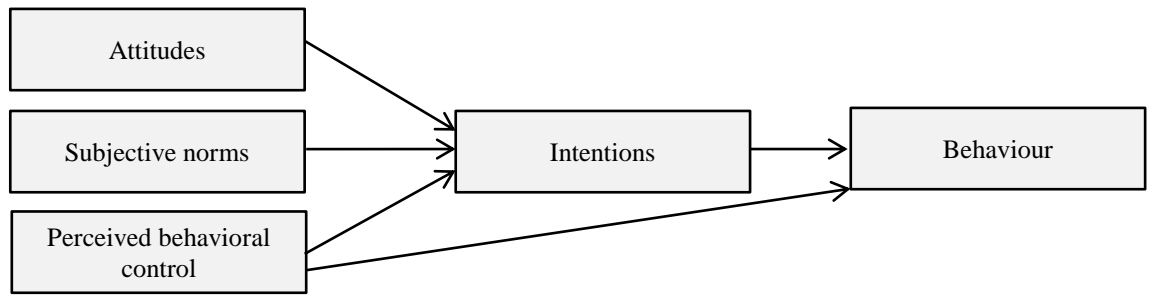

Figure 1. Relationships between the theory of planned behavior's three constructs (attitudes, subjective norms, and perceived control) and intentions and behavior. Intending to engage in a behavior is more likely when people think it will lead to positive consequences, when it has social approval, and when it is perceived to be under the individual's control. Actually engaging in the behavior is more likely when people intend to do it and when they perceive themselves as having control over the target behavior.

**Sharing and social norms**

Many ICTs let users share records of their behaviors with others, a feature that allows social norms to form. More specifically, perceiving a behavior to be widespread sets up a "descriptive" norm in favor of that behavior. Therefore, making people aware of the actions of others can mold their behavior. To give a non-ICT example, a study in the Arizona Petrified Forest found that visitors stole more petrified wood when signs emphasized the frequency of such theft compared to when signs simply discouraged stealing.[13]

Behavior sharing is a central feature of some ICT devices. For example, systems have been prototyped that aim to reduce organizational paper use by giving teams feedback on their relative printer usage[14] and to increase teenage exercise by delivering updates on the activity of group members.[15] In other cases, behavior sharing is a peripheral feature. For example, the FitBit is primarily a device for tracking personal exercise and sleep cycles, but it also includes the ability to share progress with others. Behavior sharing mechanisms are particularly valuable when they make salient behaviors that might otherwise be overlooked (such as contributions to housework[16]).

**Suggestions and perceived control**

ICTs can also foster perceived behavioral control. One way is by giving users concrete suggestions and guidance, which make desirable outcomes seem within their grasp. For example, the GoalPost and GoalLine systems provide users with suggestions for exercise activities sorted into categories (such as strength or flexibility).[17] The systems therefore make it easy for users to assemble a balanced set of achievable exercise goals. Similarly, the ShutEye system lets users specify their desired sleep schedule and then provides real-time information on when activities such as caffeine consumption, exercise, or napping are appropriate.[18] This display lets users see at a glance how to structure their behaviors to best match their desired sleep schedule. Both systems give users a sense that a desirable outcome is within their capabilities, and hence they would be expected to be particularly useful in cases where users lack this sense of control.

**The risk of undermining positive intentions**

Although often beneficial, device features such as sharing information about behavior and providing suggestions also contain the potential to undermine desired behaviors. If a system portrays the desired outcome as nonnormative or reduces personal control, then it might actually lower intentions to engage in the behavior. For example, imagine that the above systems show users that their friends use far more paper than they do, or that a healthy sleep cycle is inconsistent with their existing lifestyle. Under these circumstances, the system is less than useless; it will undermine a user's engagement in positive behaviors. Furthermore, in the context of ICT devices, waning intentions are likely to be accompanied by a drop-off in usership. People are unlikely to persist in using a device if they no longer care about changing their behavior. Hence, failing to account for these factors could harm both behavioral outcomes and user retention.

A (non-ICT) example of such a boomerang effect in action is given by P. Wesley Schultz and his colleagues, who showed that giving information on average neighborhood electricity usage decreased the amount used by those above the average, but actually increased the amount used by those below the average.[19]

Steps can be taken to mitigate against these possibilities. For example, when sharing behavior, emphasizing the evaluations people have of a behavior (the "injunctive" norm) rather than simply what they do (the "descriptive" norm) can help avoid widespread-but-disapproved-of behaviors from becoming normative. Schultz and his colleagues showed the efficacy of this approach; accompanying the descriptive feedback with a smiling or frowning emoticon eliminated the boomerang effect just described.[19] In ICT systems, this could also be achieved by letting users publically like or dislike behavior-relevant stimuli.

With regard to perceived control, it's important to ensure that devices make demands that people feel are within their capabilities. Thus, recommendations must be concrete and comprehensible, such that users understand what they need to do to act upon them. It might also be beneficial to err on the side of making overly easy recommendations to begin with, before building to harder targets.

## Self-Determination Theory

Self-determination theory embodies a humanistic approach to motivation and well-being. If the goal of a technology is to instill deep commitment to an outcome, then the theory forms a useful guide to relevant factors and potential obstacles.

**Psychological needs, rewards, and motivation**

Self-determination theory centers around two main themes. First, people have the inherent tendency to be proactive and internally motivated, but they can also become passive and indolent when their psychological needs aren't being met. Second, people have at least three psychological needs: competence (that is, feeling capable), autonomy (that is, feeling like they are in control of their own decisions), and relatedness (that is, feeling connected to people around them). When these three needs are being met, people will tend to be active and internally motivated. However, as Figure 2 illustrates,

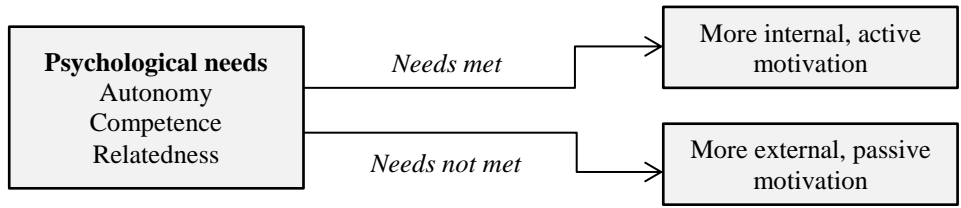

Figure 2. The relationship between need fulfillment and motivation in self-determination theory. People become more passive and more dependent on external rewards when their needs for competence, autonomy, and relatedness aren't met.

when these needs are not met, people will become more passive and dependent on external rewards.[20]

One important consequence of this theory is that overly controlling attempts to inculcate positive behavior are likely to backfire. Methods such as demanding compliance without justification, imposing goals and rewards, and threatening punishment for noncompliance all thwart the need for autonomy. Any change in response to such methods is likely to be superficial; people might be motivated to engage in the specific behavior but only to get the reward or avoid the punishment. Once rewards and punishments are removed, the desired change is likely to disappear. Furthermore, such interventions will actually damage the motivation of people who are already internally driven. People who would engage in the behavior for its own sake might come to engage in it simply for the external pay-off.

To give an example, imagine an organization that rewards its engineers for submitting proposal documents (an innovative behavior). In this case, we would expect to see an increase in the rewarded behavior. However, this newfound innovative culture might decline if the reward scheme is suspended; further, staff members might not demonstrate signs of innovation aside from turning in proposal documents, and those staff members who previously enjoyed coming up with new ideas might come to engage in this behavior simply to gain the reward. Empirically speaking, a meta-analysis of 183 experimental studies supported the contention that extrinsic rewards undermine intrinsic motivation for interesting tasks.[21]

**User autonomy and deep change**

In terms of implications for ICT design, self-determination theory suggests that the key to inducing deep, internalized commitment is to provide people with a sense of ownership in their decisions. Respecting the autonomy of individuals isn't a burdensome moral requirement with which interventions must comply; rather, supporting autonomy is an essential part of designing an effective intervention.

Methods for autonomy support vary across domains, but some examples include
- providing meaningful choices (which need not imply multiplying options— in cases with many options, it might require reducing the set to a more manageable number);

- explicitly acknowledging the anxieties felt by users; and
- providing ways for users to actively explore behavior-change strategies (rather than prescribing a linear series of steps).

Arlen Moller, Richard Ryan, and Edward Deci provide an accessible discussion of what this means in practice, as well as some concrete examples of interventions using these principles.[22] Also, the selfdeterminationtheory.org website contains an extended list of peer-reviewed empirical papers utilizing the theory.

**Games, exploration, and autonomy support**

Many ICT devices take advantage of game-like features. When these features support users' autonomy, they can be effective at promoting internal motivation.[23] One example is the Zombies, Run! application. As I described earlier, Zombies, Run! encourages jogging behavior by intertwining a runner's activities with an ongoing immersive narrative of a small community surviving a zombie outbreak. By casting users in the role of "runners" searching for supplies and avoiding zombie packs, the app reframes jogging behavior from a burdensome duty to an active, autonomous process of exploration. (A fuller discussion of the persuasive power of interactive narratives is available elsewhere.[24])

Examples that are conceptually similar, but less structured, include the Piano Stairs and Social Stairs projects.[25] Both systems modified public staircases to play noises when stepped on. The Piano Stairs played notes corresponding to a piano keyboard, whereas the Social Stairs played a richer range of sounds depending on the configuration of multiple users on the staircase. These systems let users actively explore the range of sounds that can be created, and thus tapped into their intrinsic motivation.

**Gamification features and shallow motivation**

Although game features such as narrative and exploration can promote internal motivation, other gamification features might have undesirable side effects. These features transparently seek to mold user behavior, especially through the use of punishments or incentives. Examples include points systems, rewards for task completion, and competitive leaderboards. Many devices utilize these features, including FitBit, GoalPost / GoalLine, and even Zombies, Run! itself.

As in the earlier example with the engineers, such features might increase the desired behavior, but they also foster a more external motivational stance. The risk is that any change would be dependent on the continued reward and specific to the rewarded behavior. So, for example, if a jogging app posts progress on a leaderboard, users might initially run more to climb up the ranks. However, their motivation is contingent on this reward; once they've reached a level in which they're no longer increasing in position, they might not maintain their running behavior. Furthermore, they're unlikely to manifest change in any other health behaviors— such as improving diet and sleeping more — and might stop using the application. Extrinsic motivation might not be a problem for some applications. Such features can be useful if rewards can be maintained indefinitely, the device also includes autonomy-supportive features (such as exploration or narratives), or the goal is simply to motivate a short term behavior change (such as to establish a new habit). However, self-determination theory makes it clear that these features can have undesirable side effects if used in isolation.

## Control Theory and Implementation Intentions

Control theory provides an influential account of how people translate their goals into actions.[26]

**Goals, behavior, and feedback**

As Figure 3 shows, control theory applies the idea of discrepancy-reducing feedback loops to human action; goals are taken as standards, and the process of goal pursuit is taken as an ongoing attempt to minimize discrepancies from these standards. Hence, if a person resolves to exercise more, going to the gym will lead to positive feelings (as the person moves closer to his or her desired end-state) and failing to go will lead to negative feelings (as the discrepancy isn't being reduced). Failure to make satisfactory progress motivates either increased effort or abandoning the goal, depending on how achievable success is assessed to be.

Control theory provides a clear conceptual map of the process of goal pursuit. In particular, the feedback loop captures the inherently dynamic nature of goal striving. As Paschal Sheeran and Thomas Webb show, this analysis provides an excellent framework for unifying existing findings.[27] By reducing goal pursuit to a number of component subprocesses—that is, setting goals, monitoring progress, altering responses, and disengaging from unproductive goals—control theory provides a good way to bring together seemingly unconnected empirical effects.

One behavior change technique that fits well in a control theory framework is setting implementation intentions. These are simply if–then plans for pursuing desired outcomes, specifying how a particular behavior will be triggered by cues in the environment. For example, rather than relying on a general desire to lose weight, a person might decide that if he feels hungry, then he will eat an apple, and if he can get home before 7 p.m., then he will go running. As a review of 94 independent empirical tests shows, implementation intentions are effective because they both make opportunities for action more salient and automate behavioral responses.[28] From a control

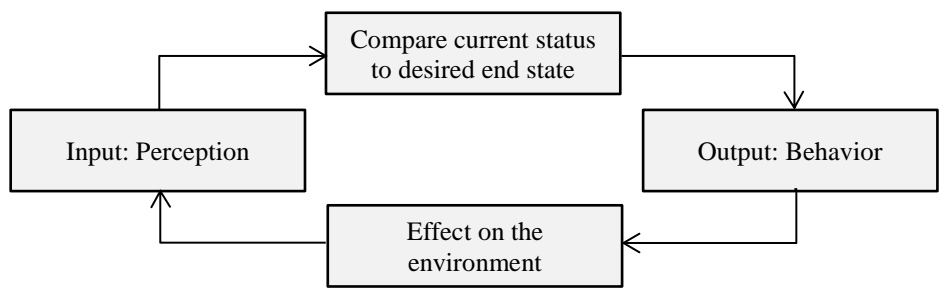

Figure 3. In control theory, goal pursuit is a discrepancy-reducing feedback loop. That is, the goal is viewed as a standard, and the process of pursuing that goal as an attempt to minimize the gap between that standard and a person's current state.

Psychological Frameworks for Persuasive ICTS | 10theory perspective, this is unsurprising; thinking about concrete action plans forces people to set measurable standards and prepare for goal-consistent action.

**Facilitating goal pursuit**

When a person isn't engaging in a desirable behavior, it's intuitive to conclude they lack the appropriate intentions. So, for example, Mike eats too much junk food and doesn't exercise. It's easy to conclude he isn't really motivated to be healthy. However, an alternate possibility is that Mike is indeed motivated to live a healthier life but is struggling to find the best way to achieve this end. On this analysis, an intervention doesn't need to provide him with the appropriate motivation, but it does need to help him clarify and act upon his goals.

As conceptualized in control theory, several elements must be in place for goals to be successfully attained. First, the goal itself must be clear and specify a desirable reference value. Setting targets works better than generally resolving to do one's best. Second, people must have the feedback information required to monitor their progress. Third, people must be able to recognize opportunities for goal-consistent action. Although this point sounds straightforward, people juggle many projects and priorities; as such, there's a real danger that they will miss opportunities for goal pursuit.

**Goal specification and tracking**

Many devices let users set goals and track progress toward them. These features are central to the GoalPost and GoalLine systems.[17] In particular, these mobile applications utilize two features. First, as I described earlier, they let users select weekly exercise activities from a range of categories, including strength and flexibility. This facilitates the setting of goals that are both concrete (that is, users know precisely what they are aiming to achieve) and well informed (that is, the goals are balanced across different categories). Second, the app provides a self-monitoring feature that lets users record their exercise behavior.

More generally, the availability of ICT has made it much easier for users to track their progress toward goals. While keeping quantitative logs of personal metrics is a longstanding practice (as in the practice of balancing checkbooks), pervasive devices have made data collection, processing, and visualization extremely easy (or, in many cases, entirely automatic):

- Sensors can monitor users' physical activity and sleep patterns (as in FitBit).
- Web applications let users track their finances (as in Mint), Internet brows ing habits (as in Voyurl), and mood (as in MoodPanda).
- Phone apps enable logs of spatial location (as in Foursquare), journal entries (as in Momento), and eating habits (as in 80 bites).

Thus, users can almost effortlessly track their progress across a whole range of domains.

**Maintaining progress to avoid boomerang effects**

Although goal setting and tracking are important, they have the potential to backfire if users consistently fail to move forward. If an ICT makes it apparent to users that they're not progressing toward their goals, it risks prompting goal disengagement. Indeed, there's some empirical evidence that people who disengage from unproductive goals are happier than those who persist in pursuing them. [29]

So, for example, if GoalPost / Goal-Line users consistently fall short of their exercise targets, they might become jaded about the possibility of living a healthier life and thus stop making efforts in this area. This disengagement might also promote user attrition. The GoalPost / GoalLine designers let users set both a primary and a secondary goal for the week, in an attempt to avoid this problem. However, another way to mitigate the risk would be to help users formulate plans for goal pursuit. As described earlier, implementation intentions provide a well-validated basis for these kinds of features. By setting appropriate situational cues for action, these behavioral plans can psychologically automate goal pursuit.

ICTs can help users form goal plans in several ways. First, apps could simply prompt users to formulate appropriate plans - that is, to not only think about the behavior they want to engage in but also the circumstances under which they will engage in it. Second, they could provide suggestions on useful if-then plans, either drawing on expert knowledge or the experiences of other users. Third, they could prompt users to track occurrences of the relevant cues, allowing them to diagnose missed opportunities for action.

Finally, devices might be able to help users recognize situational cues. The most obvious example here is time-based reminders, as implemented in many calendar and to-do apps. However, other possibilities include making behavior prompts dependent on users' location or activities, such as browsing to particular websites or using social media.

**Conclusion**

Pervasive technologies provide unprecedented opportunities for fostering change in people's thoughts, feelings, and behaviors. Webpages and apps make it easy for users to share their experiences with others. Personal sensors, always-at-hand smartphones, and cross-device applications all enable greater tracking of data about everyday behavior. Further, increased smartphone processing power and mobile broadband make external data sources ever-more accessible to users. These widely available resources can be a potent force, both enabling new methods for inculcating positive change as well as distributing them to an extremely large audience.

The aim of this article has been to show that psychological theories can assist in unlocking this potential. Specifically, the theories outlined here provide conceptual frameworks for designing ICTs that inculcate positive changes in users. Thinking about behavior in this theoretically informed way offers at
least three benefits.

First, taking a theoretically informed approach allows developers to be clearer about why their interventions should be effective. For example, a mobile app for promoting healthy eating might aim to act by
- using social norms, such as showing how widespread healthy eating is;

- supporting autonomy, by providing users with tools to make informed choices; or
- facilitating goal pursuit, by setting behavioral plans for healthier eating.

As HCI researchers have argued, this clarity about mechanism-of-action is useful for both design and evaluation of systems.[9]

Second, this approach draws attention to fertile areas for future development. In this article, for example, I've introduced the psychological concepts of perceived control, autonomy support, and implementation intentions. Although these aspects of behavior change are supported to some extent by existing technologies, a more explicit consideration of how ICTs can facilitate them in specific use cases will generate fruitful new possibilities.

Third, these theories draw attention to important ways in which interventions can backfire. People are complicated, and any attempt to change their behavior runs the risk of increasing undesirable outcomes. An awareness of psychological theories doesn't offer complete immunity against this possibility, but it can highlight important points for developers to consider. For example, if a device allows users to communicate, is there a danger of negative social norms forming? If reward mechanisms are going to be utilized, are there sufficient additional measures to ensure any change is deeply rooted? If behavior is going to be tracked over time, will the resulting data be meaningful to users? Having a theoretical backdrop allows these pitfalls to be identified and appropri-ate measures taken.

Psychology offers a fertile resource for ICT design and development, as well as decades of careful, rigorous, empirical testing and well-validated theories that address people's thoughts, feelings, and behaviors. These theories can provide developers with important insights into how technologies affect users, contributing clarity and structure at many stages of the development process.